\documentclass[twocolumn,showpacs,amsmath,amstex,amssymb,prb]{revtex4}
\usepackage{graphicx}
\usepackage{epstopdf}

\newcommand{\br}{{\bf r}}

\begin{document}
\author{Emil Prodan$^1$ and Roberto Car$^2$}
\address{$^1$Department of Physics, Yeshiva University, New York, NY 10016 and $^2$Department of Chemistry and Princeton Institute fot the Science and Technology of Materials, Princeton University, Princeton, NJ 08544}
\title{Tunneling conductance of amine linked alkyl chains}

\begin{abstract}
The tunneling transport theory developed in Phys. Rev. B {\bf 76}, 115102 (2007) is applied to molecular devices made of alkyl chains linked to gold electrodes via amine groups. Using the analytic expression of the tunneling conductance derived in our previous work, we identify the key physical quantities that characterize the conductance of these devices. By investigating the transport characteristics of three devices, containing 4, 6, and 8 methyl groups, we extract the dependence of the tunneling conductance on the chain's length, which is an exponential decay law in close agreement with recent experimental data. 

\end{abstract}

\date{\today}

\maketitle

Alkyl chains are among the simplest and first organic molecular chains considered in molecular electronics experiments. In spite of a large number of studies performed in the last decade, the quest for a thorough understanding of the transport characteristics of these devices is still open. Recent experiments by Venkataraman et al.,\cite{Venkataraman:2006lq} in which amine groups were used as links between molecular chains and gold electrodes, reported very precise measurements of the linear conductance $g$. In particular, Venkataraman and collaborators were able to measure the variation of $g$ with the number of methyl groups, $N$, with $N$ ranging from 2 to 8. As expected for tunneling transport, the results showed an exponential decay of $g$ with $N$: $g=g_ce^{-\beta N}$. The main novelty of these experiments is an unprecedented accuracy in the determination of the pre-exponential factor $g_c$. This opens the way to measuring with great precision the effect of the contacts on the transport characteristics of organic molecular devices.~\cite{Venkataraman:2006db,Cui:2001ek}

Tunneling transport is an old subject, but only recently it was formulated in a modern framework\cite{Mavropoulos:2000cr,Tomfohr:2002oq,Tomfohr:2004ve,Fagas:2004uq} in which  the tunneling resistance, and more precisely the exponential decay factor $\beta$, is extracted from the complex band structure of the molecular chain. This procedure extends far beyond the limitations of simple models that approximate electron tunneling in molecular devices using square potential barriers. The present authors contributed to this formalism by deriving an analytic expression for the contact conductance $g_c$.\cite{Prodan:2007qv} This expression gives $g_c$ as an overlap integral between three well defined and physically relevant quantities: the spectral density of the device at the Fermi level, the potential perturbation of the metallic contacts on the molecular chain, and the evanescent electron waves traversing the molecular chain. Our theory provides novel insight on the electronic structure mechanisms that underlie the experiments of Ref.~\onlinecite{Venkataraman:2006lq}. In particular, we quantify the effect of the alignment of the molecular levels with the Fermi level of the metal and the effect of the chemical bonds between the link groups and the electrodes. We find that in devices based on alkyl chains the conductance depends less sensitively than in devices based on benzene rings on the Fermi level alignment of the molecular levels. This is a consequence of the complex band structure of the alkyl chains, which is characterized by a large insulating gap. We also find that the contact conductance in the amine linked alkyl chain devices is determined to large extent by the chemical contact between a single Au atom and the amine group. Indeed the direct Au-N link contributes to more than ~60 percent of $g_c$, the adjacent layer of Au atoms contributes to less than ~30 percent of it, and the remaining Au layers contribute to the rest (less than 10 percent of $g_c$). We also give a precise quantitative assessment of the lateral extent of the region that is relevant to tunneling transport.

{\it Theoretical framework.} We consider a device consisting of a long but finite periodic molecular chain attached to infinite metallic electrodes. The chain is oriented along the $z$ axis. We assume that a self-consistent Kohn-Sham calculation for the entire device has been completed. Then, the adiabatic linear conductance is given by\cite{Prodan:2007qv}
\begin{equation}\label{adiabaticg}
g \equiv \int d \br_\bot  \int d\br'_\bot \ \sigma_{zz}^{\text{\tiny{KS}}}(\br_\bot,z,\br'_\bot, z'),
\end{equation}
where $\sigma_{zz}^{\text{\tiny{KS}}}({\bf r},{\bf r}')$ is the the $zz$ component of the Kohn-Sham conductivity tensor. The validity of this approximation was discussed in Refs.~\onlinecite{Prodan:2007qv}, \onlinecite{A.-Kamenev:2001uq}, and \onlinecite{Koentopp:2006ys}. As a consequence of charge continuity the right hand side of Eq.~\ref{adiabaticg} is independent of the location of $z$ and $z'$. In our case it is convenient to take both $z$ and $z'$ in the middle of the chain.

\begin{figure*}
  \includegraphics[width=17cm]{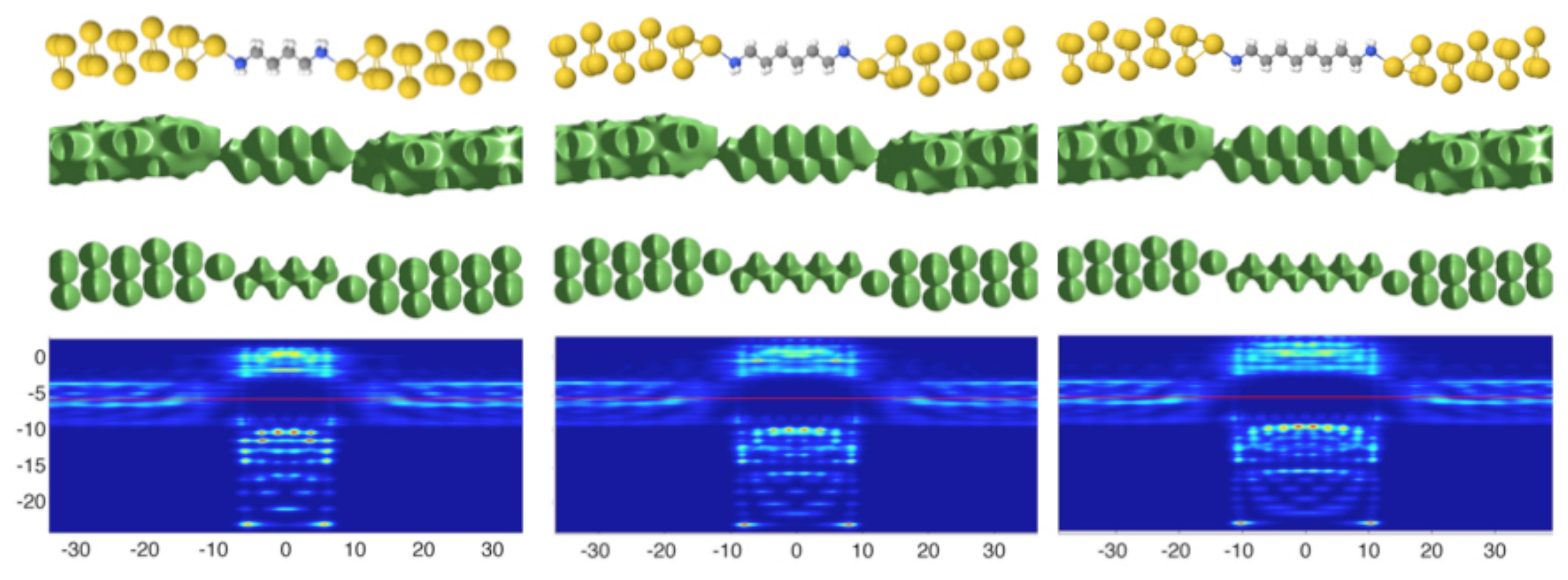}\\
  \caption{Atomic configurations of the molecular devices. Under each atomic configuration, the figure shows iso-surfaces of the self-consitent electron density (corresponding to a value which is $1\%$ of the maximum electron density), of the self-consistent Kohn-Sham potential (corresponding to a value which is $5\%$ of the maximum potential) and the planar average (over xy) of the local density of states, shown as a density plot with energy (in eV) on the vertical axis and z coordinate (in a.u.) on the horizontal axis.}
 \label{Equilibrium}
\end{figure*}  

It is also convenient to treat the device consisting of chain plus leads as a strictly periodic system strongly perturbed by the leads. The Kohn-Sham potential of the system, $V_\text{eff}({\bf r})$, is not strictly periodic inside the chain, because the effect of the leads can propagate deeply into the chain. However, one can construct a strictly periodic potential $V_0({\bf r})$, by replicating the portion of $V_{\text{eff}}({\bf r})$ that belongs to the unit cell of the chain located in the middle of the device. In the present case this cell contains two molecular CH$_2$ units. Then the total effective Hamiltonian $H=-\nabla^2 +V_{\text{eff}}({\bf r})$ can be written as (we use $\hbar$=1, $2m$=1 units)
\begin{equation}\label{deco}
H=-\nabla^2 +V_0({\bf r})+\Delta V ({\bf r})\equiv H_0 + \Delta V({\bf r}),
\end{equation}
where $\Delta V ({\bf r})=V_{\text{eff}}({\bf r})-V_0({\bf r})$.
We further decompose $\Delta V$ into left and right parts relative to the mid plane of the device: $\Delta V$=$\Delta V_\text{\tiny{L}}$+$\Delta V_\text{\tiny{R}}$. The reason for putting in evidence a periodic Hamiltonian $H_0$ in Eq.~\ref{deco} is that periodic potentials are simple and lead, in particular, to a compact expression for the Green's function.\cite{Prodan:2006yq} Moreover, Ref.~\onlinecite{Prodan:2005vn} showed that, whenever a periodic system is perturbed by a distant $\Delta V$, as in the present case, one can derive a non-perturbative expression for the total Green's function, which can be used to calculate the Kohn-Sham conductivity tensor needed in Eq.~\ref{adiabaticg}. In particular, in the limit of long chains, this approach led us to the following asymptotic expression:\cite{Prodan:2007qv}
\begin{equation}\label{insulatingg0}
g=\frac{1}{\pi} \frac{\Theta_\text{\tiny{L}} \Theta_\text{\tiny{R}}}{(\partial_k \epsilon_{k})^2}  e^{2ikL}.
\end{equation}
Here $k$ is the complex wavenumber of the evanescent Bloch solution $\psi_k({\bf r})$ of the periodic Hamiltonian $H_0$ having minimum imaginary wavenumber and energy $\epsilon_k$ equal to $\epsilon_F$, the Fermi level of the leads. The tunneling coefficient $\beta$, introduced earlier in this paper, is related to $k$ via $\beta=2\mbox{Im}[k]b$, $b$ being the lattice constant of the chain. The contact conductance $g_c$ is the pre-exponential factor in Eq.~\ref{insulatingg0}. The $\Theta_\text{\tiny{L}}$ coefficient is defined by:
\begin{equation}\label{theta1}
\begin{array}{c}
\Theta_\text{\tiny{L}}=2\pi i \int d \br \int d \br'  e^{-ik(z+z')} \times \medskip \\
u_{-k}(\br)\Delta V_\text{\tiny{L}}(\br)\rho_{\epsilon_F}(\br,\br')\Delta V_\text{\tiny{L}}(\br')u_{-k}(\br'),
\end{array}
\end{equation}
where $\br$ and $\br'$ are measured from the left end of the chain. Similarly
\begin{equation}\label{theta2}
\begin{array}{c}
\Theta_\text{\tiny{R}}=2\pi i \int d \br \int d \br'  e^{ik( z+ z')} \times  \medskip \\
u_{k}(\br)\Delta V_\text{\tiny{R}}(\br)\rho_{\epsilon_F}(\br,\br')\Delta V_\text{\tiny{R}}(\br')u_{k}(\br'),
\end{array}
\end{equation}
where $\br$ and $\br'$ are measured from the right end of the chain. In the above expressions, the evanescent waves have been factorized into exponentially and periodically varying parts, $\psi_k({\bf r})=e^{ik z}u_k({\bf r})$. $\rho_{\epsilon}$ is the spectral operator, $\rho_{\epsilon}=\frac{1}{2\pi i}  (G_{\epsilon+i\delta}-G_{\epsilon-i\delta } )$. Its diagonal part gives the local density of states $\rho_\epsilon(x,y,z)$ of the device. The denominator in 
Eq.~\ref{insulatingg0} is the $k$-derivative of the band energy $\epsilon_k$ evaluated at $\epsilon_F$.    

{\it Geometrical models and technical details.} We studied three devices containing 4, 6 and 8 methyl groups, linked to gold electrodes via amine groups. In the following, these three devices will be referred to as (a), (b) and (c), respectively. The corresponding atomic configurations are shown in the first row of Fig.~\ref{Equilibrium}. The alkyl chain has the same geometry as in Refs.~\onlinecite{Tomfohr:2002oq} and \onlinecite{Picaud:2003qf}. The amine groups at the two ends of the chain simply replace a methyl group, and we neglect the small difference in length between the NH and the CH bonds. Indicating by A, B, and C the stacking planes in the (111) direction of fcc gold, a device consisting of an alkyl chain and two gold leads is represented schematically by:
\begin{equation}
\text{BCBCBA-NH$_2$-(CH$_2$)$_N$-NH$_2$-ACBCBC}
\end{equation}
As shown in Fig.~\ref{Equilibrium} only one Au atom of the A plane is included, whereas each of the B and C planes is represented by three Au atoms.  The Au-N bond length is fixed to 2.4 A, and the Au-N-C angle is set to $109.5^o$. The two leads are tilted relative to the alkyl chain in order to enforce the above geometrical constraints and to permit periodic boundary conditions along the z direction between the left and right ends of the device. Our calculations were performed on periodic supercells containing the alkyl chain and the Au wires that represent the leads in Fig.~\ref{Equilibrium}. The supercell dimensions (in a.u.) for the three devices were, respectively: (a) 22.30 $\times$ 22.30 $\times$ 68.94, (b) 22.30 $\times$ 22.30 $\times$ 73.70, (c) 22.30 $\times$ 22.30 $\times$ 78.45. Periodic boundary conditions were applied in all three dimensions, but in this setup periodicity along x and y is only a matter of numerical convenience because the leads (and the molecules) are well separated from their periodic replicae in the xy plane. 

The equilibrium self-consistent Kohn-Sham calculations were performed with a real space pseudopotential code based on finite differences. We adopted a 5-point finite difference approximation for the kinetic energy operator, and used a uniform space grid with a one dimensional spacing of 0.3565 a.u., sufficient for a good convergence of the electronic band structure. This grid is commensurate with the unit cell of the periodic alkyl chain, which is the reference system in our transport calculations. We adopted the Local Density Approximation (LDA) for exchange and correlation using the Perdew-Zunger (PZ)\cite{Perdew:1981gb} interpolation of the numerical electron-gas data of Ceperley and Alder.\cite{Ceperley:1980eu} 

We used Troullier-Martin norm-conserving pseudo-potentials \cite{Troullier:1991ys} for all the atomic species. The pseudopotentials for C and N atoms had distinct {\it s} and {\it p} components and we took the {\it p} pseudo-potential as the local reference. Purely local pseudopotentials were used for the H and Au atoms. In the latter case only the outermost {\it s} electrons were treated explicitly, but non-linear core corrections were included.\cite{Louie:1982hl} This approach overestimates the work function of gold yielding values between 6.28 and 6.70 eV depending on the surface orientation.\cite{Fall:2000kx} On the other hand, when the {\it d} electrons are treated explicitly the LDA yields workfunctions very close to experiment.\cite{Fall:2000kx} In our calculation we find a vertical ionization potential of 6.4 eV for the Au wires, which should be compared to an average experimental value of $~5.4$ eV for the workfunction of the Au (111) surface.\cite{Hansson:1978ij} We will comment later on the possible consequences of an error of $~1$ eV in the level alignment of our devices. 

\begin{figure}
  \includegraphics[width=8.6cm]{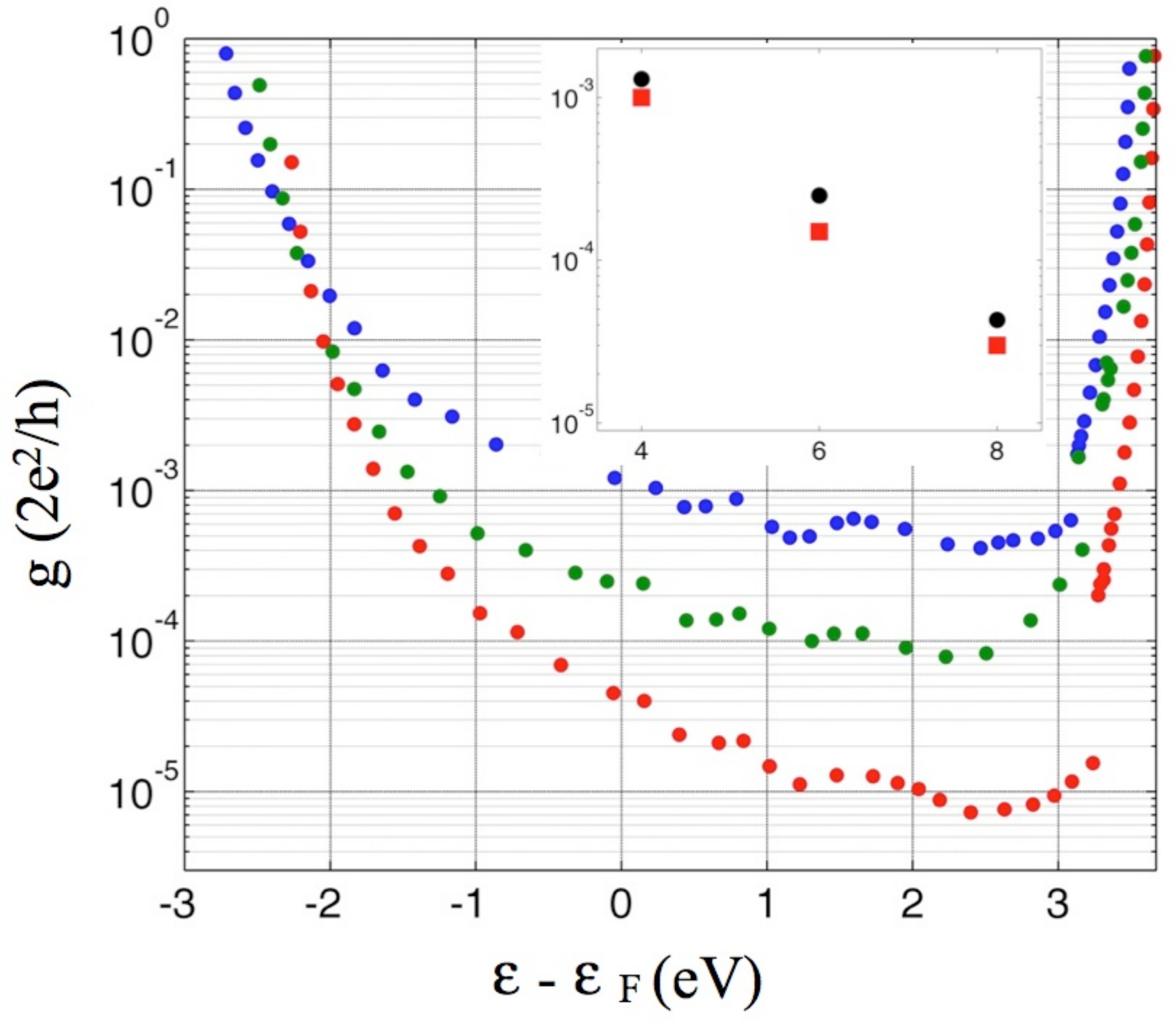}\\
  \caption{Conductance of the three devices [blue for device (a), green for (b) and red for (c)] as a function of the bias potential ($\epsilon$-$\epsilon_F$). The inset shows a comparison between our prediction (black dots) for the linear conductance and the experimental data of Ref.~\onlinecite{Venkataraman:2006lq} (red squares). The conductance values that are compared to experiment were extracted from the main plot at $\epsilon$-$\epsilon_F$=0. }.
 \label{gVn}
\end{figure}

Our real-space code yields band structures in good agreement with standard plane wave calculations. In particular, the valence band edges of the infinite, isolated alkyl chain are located at -6.4 and -20.0 eV, respectively, and the insulating gap is of 5.8 eV, in good agreement with Refs. \onlinecite{Picaud:2003qf,Montanari:1997la}.

In Fig.~\ref{Equilibrium}, we show the iso-surface of the self-consistent electron density $n({\bf r})$ corresponding to a value which is $1\%$ of the maximum electron density. We also show the iso-surface of the self-consistent potential $V_\text{eff}({\bf r})$ corresponding to a value which is $5\%$ of the maximum potential. Here $V_\text{eff}({\bf r})$ includes Hartree, exchange-correlation, and local pseudo-potential contributions. Interestingly, there are no visible differences on both iso-surfaces as we move from the CH$_2$ units to the NH$_2$ end groups. Notice that to a very good approximation both $n({\bf r})$ and $V_\text{eff}({\bf r})$ are periodic inside the molecular chains all the way to the  first atoms of the Au contacts. 

The lower panels in Fig.~\ref{Equilibrium} show the local density of states, averaged in the xy plane: $\rho_\text{av}(z,\epsilon)=\int \rho_\epsilon(x,y,z)dxdy$. The plots give a color map of $\rho_\text{av}(z,\epsilon)$ in the plane of energy $\epsilon$ and of position $z$. The Fermi level is indicated by the red line. One sees that the conducting states of the leads decay rapidly to zero inside the alkyl chain, where a spectral gap is visible. The alignment of the gap edges relative to the Fermi level is approximately but not exactly the same for the three devices. One can detect a slight band bending near the contacts, in spite of the fact that $V_\text{eff}({\bf r})$ appears to be approximately periodic all the way to the first atoms of the Au electrodes. When the energy $\epsilon$ is inside the spectral gap of the chain, $\rho_\text{av}(z,\epsilon)$ does not show any special features near the contacts. The gap is clean all the way to the first gold atoms of the electrodes, indicating that no bonding states are present.  

\begin{figure}
  \includegraphics[width=8.6cm]{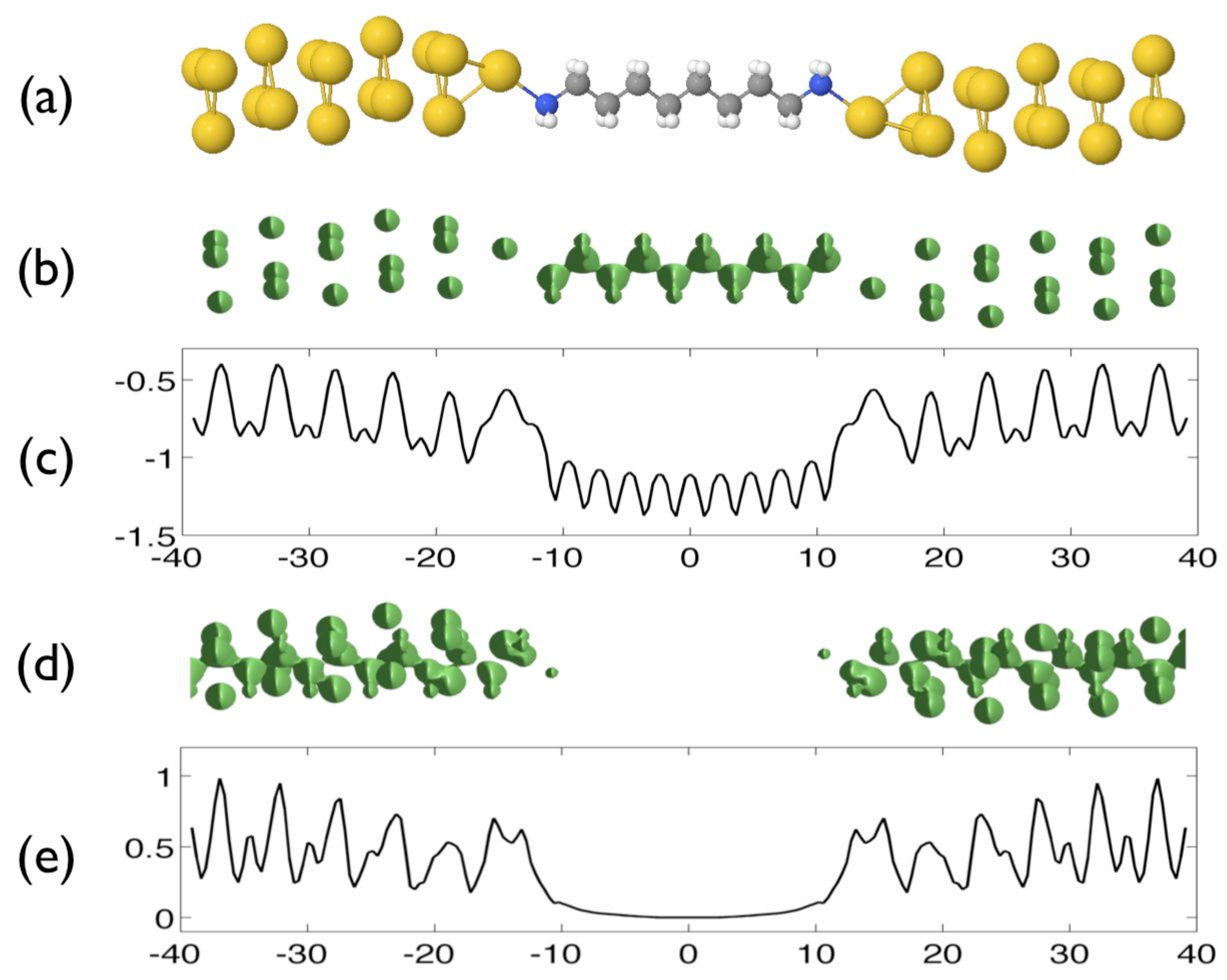}\\
  \caption{a) Atomic configuration of device (c). b) An iso-surface of $V_{\text{eff}}$. c) Planar average of $V_{\text{eff}}$ (with respect to the xy coordinates). d) An iso-surface of $\Delta V$. e) Planar average of $\Delta V$ (with respect to the xy coordinates). The energy units are Ry.}
 \label{DiffPot}
\end{figure}

{\it Band alignment.} It is important to assess the accuracy of the LDA and of the available semi-local DFT approximations in the description of the level alignment at metal-molecule interfaces.\cite{Neaton:2006le,Ouek:2006tx}  When modeling tunneling through small gap insulators, where the complex bands are parabolic, and thus depend sensitively on the energy, a relatively small change in the level alignment can affect significantly the theoretical predictions. For example, in Ref.~\onlinecite{Ouek:2007xq} which used a Generalized Gradient Approximation (GGA) treatment, the linear conductance of a device consisting of one benzene molecule linked to gold electrodes via amine groups was found to be 7 times larger than the measured experimental value, but it was shown that a rigid shift of the Fermi level by 0.5 eV would be sufficient to restore an almost perfect agreement between theory and experiment. A different situation occurs in the case of the alkyl chains, which have a large insulating gap. The relevant complex band connecting the top of the valence bands to the bottom of the conduction bands is not parabolic but becomes rapidly flat as we move away from the band edges. This means that $\beta$ is largely independent of the position of the Fermi level, as long as the latter falls sufficiently far from the band edges. In other words, good agreement between theory and experiment could occur for the conductance in spite of the DFT errors in the band alignment. Notice, however, that the connection between band alignment and conductance, although instructive, should not be taken too strictly. In fact, the Kohn-Sham levels are not quasi-particle levels, i.e. they do not correspond rigorously to observable properties. On the other hand the linear adiabatic conductance of Eq.~\ref{adiabaticg} is a static response function which is well within the realm of ground-state DFT.  

\begin{figure}
  \includegraphics[width=7cm]{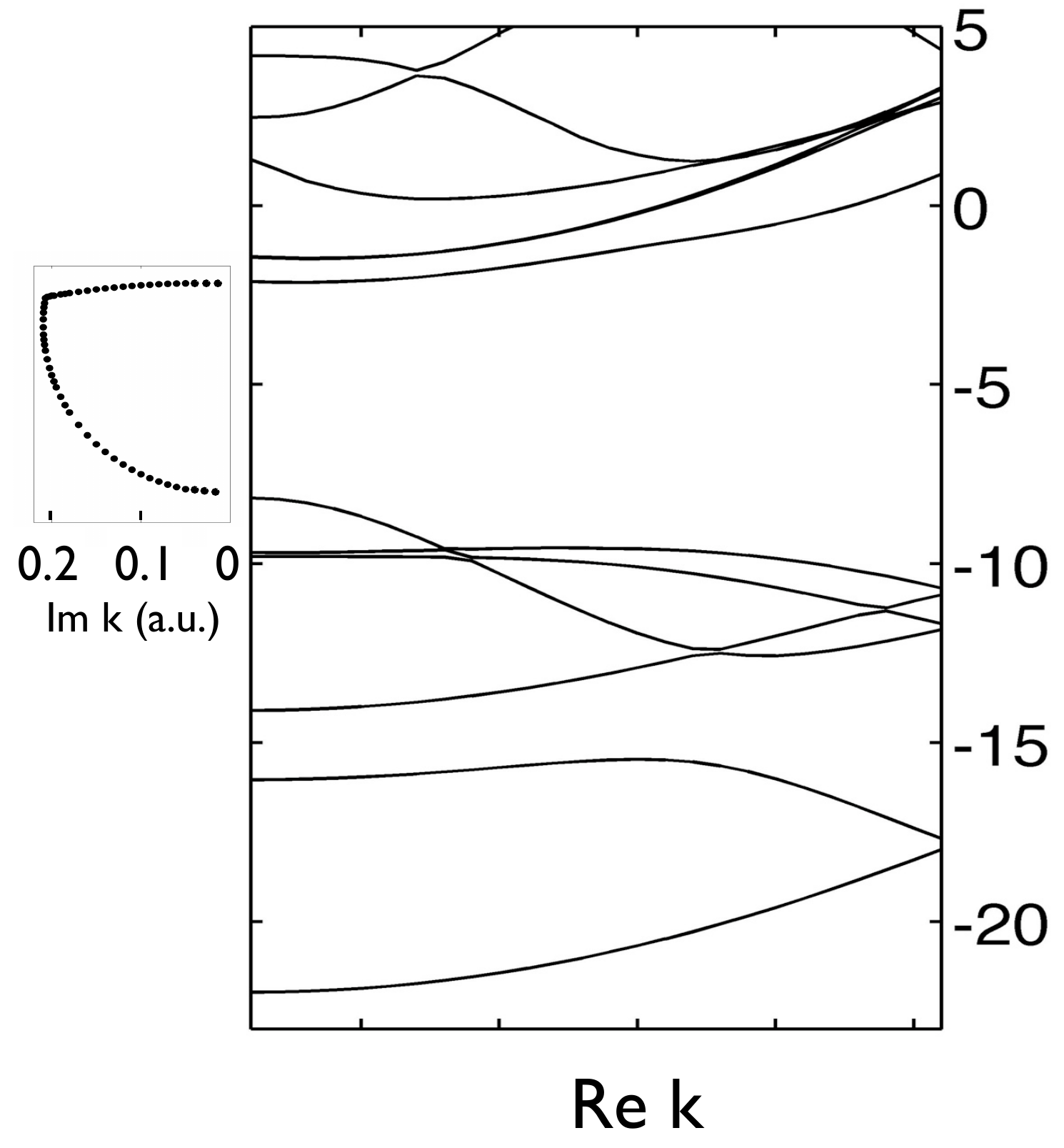}\\
  \caption{Real and complex band structures corresponding to the periodic potential $V_0$ for device (c). The main difference between these band structures and those of an infinite periodic chain is a rigid energy shift (see text). Only the complex bands with smallest Im[$k$] are shown. The energy units are eV.}
 \label{bands}
\end{figure}

Tight-binding calculations on thiol linked alkyl chains reported in Ref.~\onlinecite{Tomfohr:2002oq} give a Fermi level which is pinned at the branch point. In our calculations, the branch point (see Fig.~\ref{bands}) is very close to the vacuum level and, consequently, the scattering states (which are not included in the tight-binding treatment of Ref.~\onlinecite{Tomfohr:2002oq}) push the Fermi level down, away from the branch point, by about 2 eV. As a result the Fermi level is at ~ 3.0 eV above the valence band edge of the alkyl chain. This value reflects in part the fact that we did not explicitly include the {\it d} orbitals of gold in our calculations. When this is done, as in separate calculations performed with the plane wave pseudopotential code PWSCF,\cite{Baroni:fp} the position of the Fermi level moves to ~3.7 eV above the valence band edge of the alkyl chain.\cite{Wang:eu} Since in that energy range the complex band is relatively flat, a more accurate band structure calculation, with {\it d} electrons included, would have only a minor impact on the conductance. In fact, according to the complex band structure reported in Fig.~\ref{bands} an upward energy shift of the Fermi level by ~0.7 eV would reduce the conductance slightly, further improving the agreement between theory and experiment.

Our most important observation concerning band alignment is that the exponential decay law of the conductance with the chain's length is only valid if the band alignment does not change with the length of the chain, because otherwise $\beta$ would become a function of the chain length. The experiments reported in Ref.~\onlinecite{Venkataraman:2006lq} refer to single chains. For an isolated chain, as opposed to a dense monolayer of molecular chains connected on both sides to planar electrodes, there is no a-priori reason for the alignment to be independent of the chain's length. This is because a planar sheet of contact dipoles could be present in the case of a dense monolayer and the electrostatic effect of a planar array of dipoles would extend uniformly to infinite distance from the electrode surface, whereas the effect of a single isolated dipole would die off with distance.  In our calculations (see Fig.~\ref{gVn}), we observe a small drift of the Fermi level as the chain's length is increased. Due to the weak energy dependence of $\beta$ this small drift does not affect appreciably the exponential law, at least for the 3 chains considered here. 

{\it Conductance: Numerical results.} We evaluated the conductance according to Eq.~\ref{insulatingg0} as the Fermi level position was varied within the insulating gap. In these calculations the Fermi level was shifted rigidly, i.e. non-selfconsistently. The calculated conductance is reported in Fig.~\ref{gVn} as a function of the shifted Fermi energy $\epsilon$. In the inset the linear conductance of the three devices, at the unshifted Fermi energy, is compared to the experimental values taken from Ref.~\onlinecite{Venkataraman:2006lq}. 

In the following we explain in detail how the conductance was calculated, as it is useful to gain insight on the physics behind the results of Fig.~\ref{gVn}. In Fig.~\ref{DiffPot} we illustrate the decomposition of $V_\text{eff}$ in terms of $V_0+\Delta V$ for device (c). The second and third rows of the figure show that, up to very small deviations, the self-consistent potential is periodic inside the chain. This is confirmed by the isosurface and the xy average plots of $|\Delta V|$, which both indicate that $|\Delta V|$ rapidly decreases as we move away from the contacts inside the chain. Similar results were found for the other two devices.

 \begin{figure}
  \includegraphics[width=8.6cm]{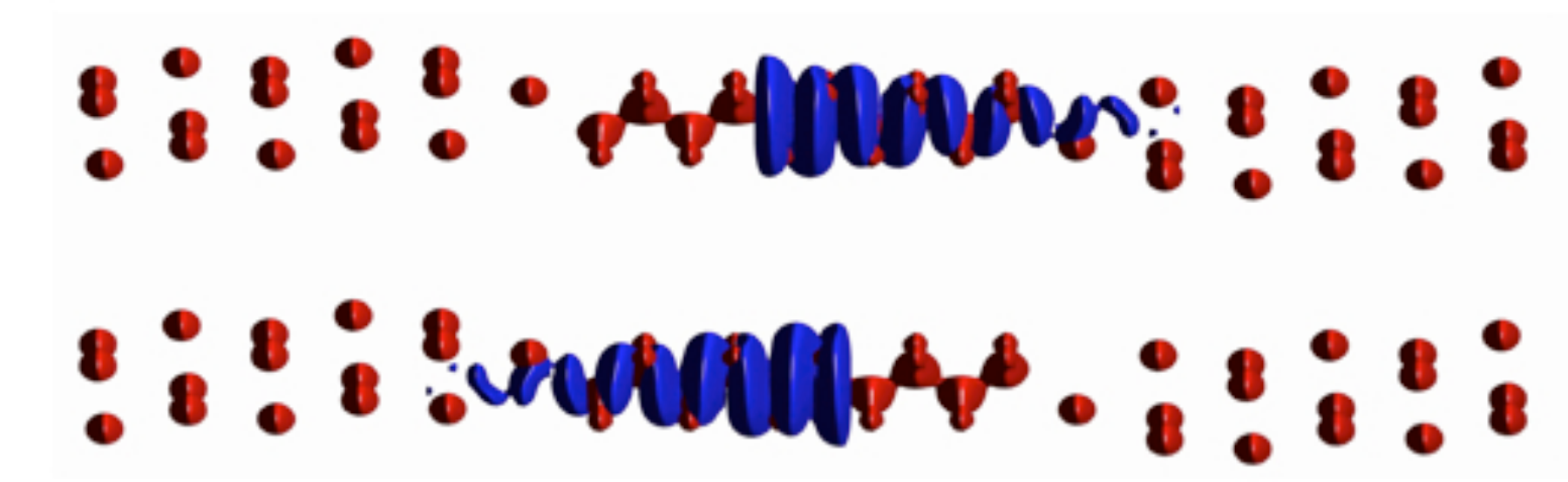}\\
  \caption{An iso-surface plot (blue surface) corresponding to 5\% of the maximum value of the evanescent Bloch functions $|\psi_{\pm k}({\bf r})|$ at the Fermi energy for device (c). For illustrative convenience the Bloch functions $|\psi_{\pm k}({\bf r})|$ were truncated at their left/right end, respectively. For reference, we also show an iso-surface plot of the effective potential (in red). }
 \label{bloch}
\end{figure}

The complex band structure corresponding to $V_0$ varies slightly when different devices are considered, but overall the bands are similar to the bands of an isolated periodic chain reported in Ref.~\onlinecite{Picaud:2003qf}. This suggests that the main difference between $V_0$ and the self-consistent potential of an isolated infinite chain is a constant energy shift. Given the complex band structure of the alkyl chains, the tunneling conductance is determined by just one complex band, the one with the smallest Im[$k$], as discussed in Refs.~\onlinecite{Tomfohr:2002oq} and \onlinecite{ Picaud:2003qf}. This complex band is shown in Fig.~\ref{bands}. It was obtained by varying continuously Im[$k$] from 0 to its maximum value, while keeping Re[$k$]$=0$. For each complex value of $k$, the spectrum of the $k$ dependent Hamiltonian:

\begin{equation}
H_k=-(\nabla-ik{\bf e}_z)^2+V_0+e^{-ik(z-z')}V_{non-local}({\bf r},{\bf r}'),
\end{equation}
with periodic boundary conditions at $z=\pm b/2$, was calculated and its eigenvalues ordered according to their real parts: Re[$\epsilon_{1k}$]$<$Re[$\epsilon_{2k}$]$<$ \ldots . We focus, in particular, on the sixth and seventh eigenvalues $\epsilon_{6k}$ and $\epsilon_{7k}$ (which take real values, see Fig.~\ref{bands}) and their corresponding evanescent Bloch functions $\psi_{6k}$ and $\psi_{7k}$. When Im[$k$]=0, $\epsilon_{6k}$ and $\epsilon_{7k}$ coincide, respectively, with the top of the valence bands and with the bottom of the conduction bands of $V_0$. For small Im[$k$], $\epsilon_{7k}$ originates from vacuum scattering states, but its character changes when Im[$k$] exceeds a certain value (see the sharp change in the complex band shown in Fig.~4)  and becomes the same as that of $\epsilon_{6k}$. 

By further increasing Im[$k$], the two eigenvalues move towards each other and coincide when $k$ reaches the branch point. The branch point is located at Im[$k$]=0.21 (a.u.), in good agreement with the finding of  Ref.~\onlinecite{Picaud:2003qf} for isolated infinite alkyl chains. 
 
At different values of Im[$k$] we evaluated Eq.~\ref{insulatingg0} for both $\epsilon=\epsilon_{6k}$ and $\epsilon=\epsilon_{7k}$ using the corresponding evanescent Bloch functions $\psi_{6k}$ and $\psi_{7k}$ to compute the $\Theta$ coefficients. The evanescent Bloch functions were also used to compute the derivative $\partial_k \epsilon_k$, making use of a Wronskian identity derived in Ref.~\onlinecite{Prodan:2007qv}. The spectral kernel was computed directly from the Kohn-Sham orbitals of the full device, $\phi_i$, and their corresponding energies $\epsilon_i$:
\begin{equation}\label{spectral}
\rho_{\epsilon}(\br,\br')=\sum_i \phi_i^*(\br)\phi_i(\br') \delta(\epsilon-\epsilon_i),
\end{equation}
with either $\epsilon=\epsilon_{6k}$ or $\epsilon=\epsilon_{7k}$.

The Dirac-delta function was approximated by $\delta(x)=x/(x^2+\delta_0^2)$, with $\delta_0$=0.1 eV.  Convergence with the number of Au layers in the leads was checked by repeating the calculations (including the self-consistent part) for devices containing from 2 up to a maximum of 5 Au layers. We found that the results are already converged for leads containing 3 Au layers.

\begin{figure}
  \includegraphics[width=8.6cm]{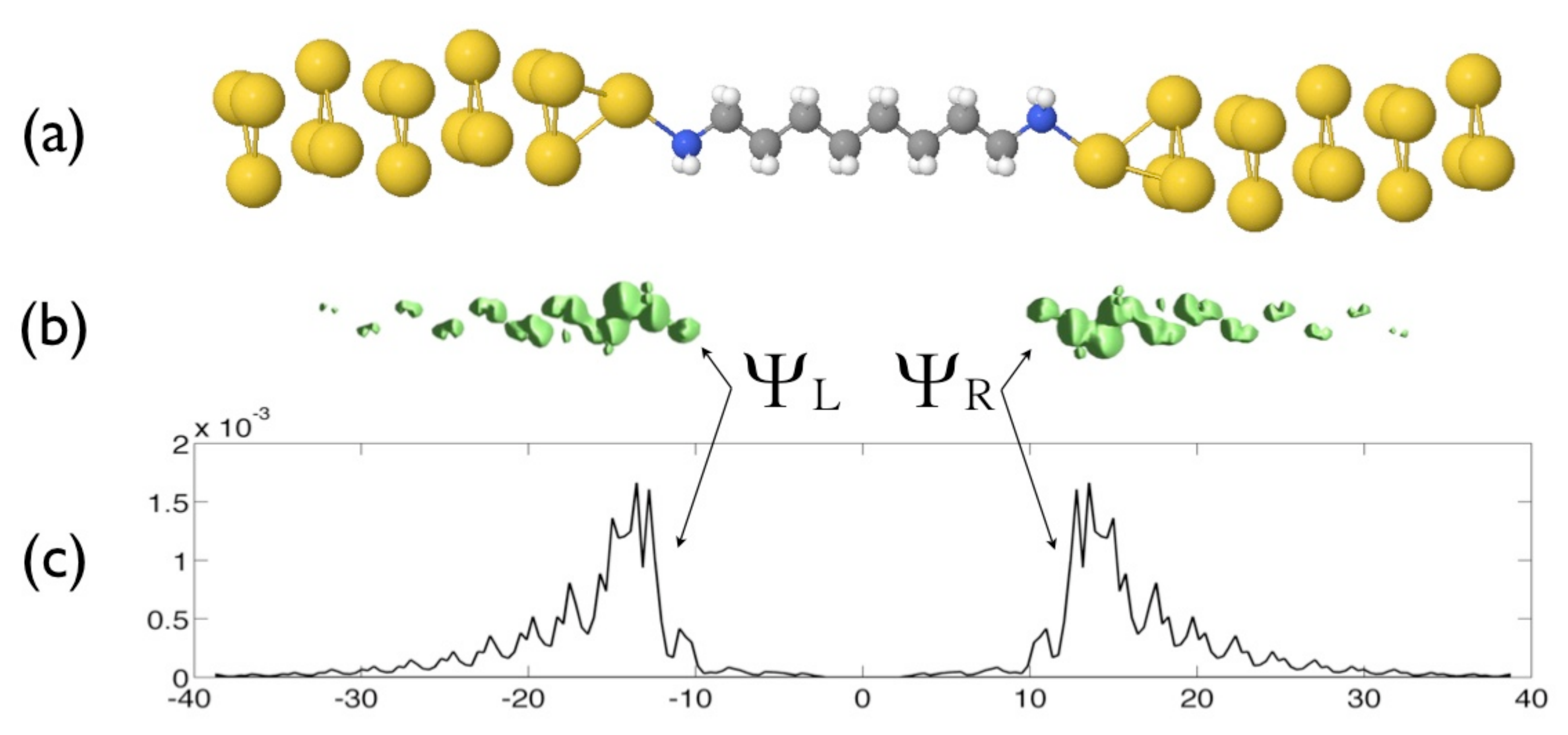}\\
  \caption{(a) Molecular device (c). (b) An iso-surface plot of $|\Psi_\text{\tiny{L/R}}({\bf r})|$, corresponding to 1\% of the maximum value of $|\Psi_\text{\tiny{L/R}}({\bf r})|$. (c) Planar average of $|\Psi_\text{\tiny{L/R}}({\bf r})|$ (with respect to the xy coordinates). The three graphs are aligned.}
 \label{Psi}
\end{figure}

{\it Discussion.} It is very instructive to plot the physical quantities that enter the definition of the $\Theta$ coefficients (see Eqs.~\ref{theta1} and \ref{theta2}). A plot of $\Delta V$ was already shown in Fig.~\ref{DiffPot} and a plot of the local density of states (i.e. the diagonal part of the spectral operator) was given in Fig.~\ref{Equilibrium}.

Fig.~\ref{Psi} shows a plot of the evanescent Bloch solutions of the periodic Hamiltonian with potential $V_0$, evaluated at the Fermi level, for device (c). 
Since the contact conductance involves an overlap of this functions with other physical quantities, plots like the ones shown in Fig.~\ref{Psi} can be used to quantitatively assess the lateral size of the contact region that is relevant for tunneling transport.  For example, the iso-surfaces in Fig.~\ref{Psi} indicate that at least 95\% of the evanescent Bloch functions are completely contained in a region narrower than the lateral size of the gold wires used in our calculation.

A crucial factor in our transport calculation, the overlap between the evanescent Bloch function $\psi_{\mp k}({\bf r})$ and $\Delta V_\text{\tiny{L/R}}$,  
\begin{equation}
\Psi_\text{\tiny{L/R}}({\bf r}) = \psi_{\mp k}(\br)\Delta V_\text{\tiny{L/R}}(\br)
\end{equation}
is exponentially localized at the left/right contacts. This means that in order to compute the conductance we only need to compute the spectral operator in Eq.~\ref{spectral} near the contacts. A plot of $\Psi_\text{\tiny{L/R}}({\bf r})$ for device (c) is shown in Fig.~\ref{Psi}. This plot allows us to quantify how the different Au layers contribute to the conductance. From panel (c) of Fig.~\ref{Psi} we extract that about 62\% of $g_c$ comes from the region occupied by the first Au atom of the leads, i.e. the contact Au atom. The adjacent Au layer contributes to only 27\% of $g_c$ and the remaining Au layers contribute to less than 10\% of $g_c$. We thus infer that variations in the contact geometry changing a little the Au-N bond length and the Au-N-C bond angle could lead to variations of $g_c$ that should not be larger than 50\% of the value found in the present calculations.
 
In conclusion, we have presented a novel approach to calculate efficiently the tunneling conductance. This scheme opens the way for first principles calculations of the conductance in devices made of long molecular chains, like e.g. the alkyl chains in the experiments of Ref.~\onlinecite{A.-Salomon:2005kx}. Our approach puts in evidence the exponential dependence of the conductance on the molecular length and links the decay length to a precise property of the complex band structure of a suitably defined periodic molecular chain. Moreover the formula for the contact conductance, i.e. the pre-exponential factor in the conductance, is relatively simple and involves overlap integrals between the evanscent waves of the periodic molecular chain and physical quantities that can be easily extracted from an equilibrium self-consistent calculation for the full device, including the electrodes and the molecular chain that connects them. Since only the region near the contacts is important, a conductance calculation can be performed on a finite model of the device, which can be conveniently done with a supercell geometry like in standard band structure calculations. Finally, the formula is simple enough to allow for semi-quantitative estimates of the conductance without the need for numerical calculations, like we did to estimate the effect on the conductance of small changes in the atomic geometry at the contacts.

For alkyl chains linked to gold electrodes via amine groups, our theoretical prediction for the tunneling conductance is in very good agreement with the recent experimental measurements reported in Ref.~\onlinecite{Venkataraman:2006lq} and the theoretical predictions of Ref.~\onlinecite{Fagas:2007fk}. We found that the level alignment in these devices is less important than one could anticipate due to the flattening of the relevant complex band away from the gap edges. Finally, we found that the contact conductance is determined mainly by the chemical link between a single atom of each gold electrode and the amine group at the corresponding end of the molecular chain. 

{\it Acknowledgments:} Partial support for this work was provided by the 
NSF-MRSEC program through the Princeton Center for 
Complex Materials (PCCM), grant DMR 0213706, and 
by DOE through grant DE-FG02-05ER46201. E.P acknowledges additional  financial support from Yeshiva University.

\end{document}